\RequirePackage{lineno} 
\documentclass[notitlepage, twocolumn,aps, pra, floatfix, superscriptaddress]{revtex4-1} 
\usepackage{amsmath}
\usepackage{amssymb}
\usepackage{graphicx}
\usepackage{color}
\usepackage{physics}
\usepackage{gensymb}
\usepackage{todonotes}
\usepackage{float} %For fig. positioning
\usepackage{xcolor} %For fig. positioning
\usepackage{cancel}%Antonis. During the drafts, to keep track of previous phrasings.

\pdfoutput=1

\newcommand{\eq}{\begin{equation}}
\newcommand{\eeq}{\end{equation}}

\begin{document}

\title{Probing many-body localization on a noisy quantum computer}
% title suggests the computer will be built soon, theory paper

\author {D. Zhu}
\email{daiwei@terpmail.umd.edu}
\affiliation{Joint Quantum Institute and Department of Physics, University of Maryland, College Park, MD 20742  USA}
\affiliation{Center for Quantum Information and Computer Science, University of Maryland, College Park, MD 20742  USA}
\author{S. Johri}
\thanks{Current affiliation: IonQ Inc.}
\affiliation{Intel Labs, Intel corporation, Hillsboro, OR 97124  USA}
\author{N. H. Nguyen}
\affiliation{Joint Quantum Institute and Department of Physics, University of Maryland, College Park, MD 20742  USA}
\author{C. Huerta Alderete}
\affiliation{Joint Quantum Institute and Department of Physics, University of Maryland, College Park, MD 20742  USA}
\affiliation{Instituto Nacional de Astrof\'{i}sica, \'{O}ptica y Electr\'{o}nica, Sta. Ma. Tonantzintla, Puebla 72840, Mexico}
\author{K. A. Landsman}
\affiliation{Joint Quantum Institute and Department of Physics, University of Maryland, College Park, MD 20742  USA}
\affiliation{Center for Quantum Information and Computer Science, University of Maryland, College Park, MD 20742  USA}
\author{N. M. Linke}
\affiliation{Joint Quantum Institute and Department of Physics, University of Maryland, College Park, MD 20742  USA}
\author{C. Monroe}
\affiliation{Joint Quantum Institute and Department of Physics, University of Maryland, College Park, MD 20742  USA}
\affiliation{Center for Quantum Information and Computer Science, University of Maryland, College Park, MD 20742  USA}
\author{A. Y. Matsuura}
\affiliation{Intel Labs, Intel corporation, Hillsboro, OR 97124  USA}

\date{\today}

\begin{abstract}
A disordered system of interacting particles exhibits localized behavior when the disorder is large compared to the interaction strength. Studying this phenomenon on a quantum computer without error correction is challenging because even weak coupling to a thermal environment destroys most signatures of localization. Fortunately, spectral functions of local operators are known to contain features that can survive the presence of noise. In these spectra, discrete peaks and a soft gap at low frequencies compared to the thermal phase indicate localization. Here, we present the computation of spectral functions on a trapped-ion quantum computer for a one-dimensional Heisenberg model with disorder. Further, we design an error-mitigation technique which is effective at removing the noise from the measurement allowing clear signatures of localization to emerge as the disorder increases. Thus, we show that spectral functions can serve as a robust and scalable diagnostic of many-body localization on the current generation of quantum computers.
\end{abstract}

\pacs{}% insert suggested PACS numbers in braces on next line

\maketitle %\maketitle must follow title, authors, abstract and \pacs


%\section{Introduction}
%Quantum computers built on a variety of platforms are increasingly being applied to the study of various quantum models with strong correlations. In this context, universal quantum computers are of particular importance because they can be used to simulate Hamiltonians, which are intractable on classical computers. A viable algorithm for a near-term quantum computer thus should be scalable, that is take time polynomial in the system size, and be robust to noise, typically achieved by designing an accompanying error mitigation technique.

Many-body localization (MBL) is a phenomenon which emerges in quantum systems with both interactions and disorder. At large values of disorder, a many-body system can fail to thermalize even at high temperatures causing it to exhibit properties like long-term memory retention, logarithmic entanglement growth in time, and area-law entanglement scaling  \cite{mbl_review_1, mbl_review_2}. The many-body localization-delocalization transition, which occurs at a critical disorder strength, is a dynamical phase transition. This necessitates the study of excited states, rather than just the ground state of the system. The study of this phenomenon in spin systems via full diagonalization exhausts classical computational power for a system of about 20 spins \cite{alet_review}. Specialized approximate schemes such as tensor network methods can in principle handle larger system sizes but tend to only work well for short-range interacting systems in one-dimension away from the phase transition  \cite{tensor1, tensor2}. Many open questions still abound  regarding the effects of symmetry, topology, dimensionality, long-range interactions, thermal inclusions, and the universality class of the disorder, especially near the phase transition. Better simulations of this phenomenon would also lead to a deeper understanding of fundamental concepts in quantum thermodynamics such as the eigenstate thermalization hypothesis. Thus, the study of a many-body localized system has been proposed as a benchmark for showing the utility of near-term quantum computers \cite{Childs9456}. %A viable algorithm for probing MBL on a near-term quantum computer should be scalable, that is take time polynomial in the system size, and be robust to noise, a property that may be achieved by designing an accompanying error mitigation technique.

Experimental efforts to probe MBL include quantum simulators consisting of thousands of cold atoms \cite{bloch1,bloch2,bloch3,bloch4,bloch5, bloch6} and a Hamiltonian whose disorder arises from the superposition of lattice potentials with incommensurate wavelengths. Another set of leading examples are experiments on trapped ions with tens of spins, which investigate the role of disorder in long-range Ising chains \cite{Smith2016, hess2017non}. Finally, up to three interacting photons in an array of transmons with random on-site energies have been studied \cite{martinis1, martinis2}. A limitation of all of these experiments is that they are specialized to a particular class of Hamiltonians that are native to the system and therefore cannot address many open questions about MBL. The only simulation of MBL on a quantum computer operated in a universal fashion was limited to a 2-spin system realized with transmon qubits \cite{zou2020}. Additionally, the energy statistics and entanglement entropy studied in \cite{martinis1, martinis2} take exponentially longer to measure as the number of interacting particles increases. Another problem arises from the noise in near-term quantum computers, which manifests itself as a thermal bath coupled to the system. Since diagnostics like level statistics and entanglement growth have been shown to revert to thermal behavior on even weak coupling to a thermal bath \cite{johri_mbl}, they are particularly unsuitable for the study of localization on such near-term devices. 
%Another proposal to use phase estimation to prepare MBL states within an energy window is unsuitable for near-term quantum computers since it requires very deep circuits \cite{bauer}.

Here, we introduce a technique for studying MBL on universal quantum computers by measuring the spectral functions of local operators. These carry signatures of localization that are known to survive coupling to a thermal bath as long as it is weaker than the characteristic energy scales of the model \cite{johri_mbl}. We measure spectra for the Heisenberg model with disordered magnetic fields along two directions implemented by 3 qubits on an ion trap quantum computer. In the many-body localized phase, the spectral functions exhibit a discrete nature, and after averaging over disorder, display a suppression of amplitude or "soft gap" at low frequencies, compared to the thermalized phase. In addition to the natural robustness to noise of our chosen observables, we also design an error mitigation scheme specific to the study of disorder-averaged spectral functions. %, which additionally removes almost all errors due to imperfect quantum operations.

\begin{figure*}
\centering
 \includegraphics[width=0.75\textwidth]{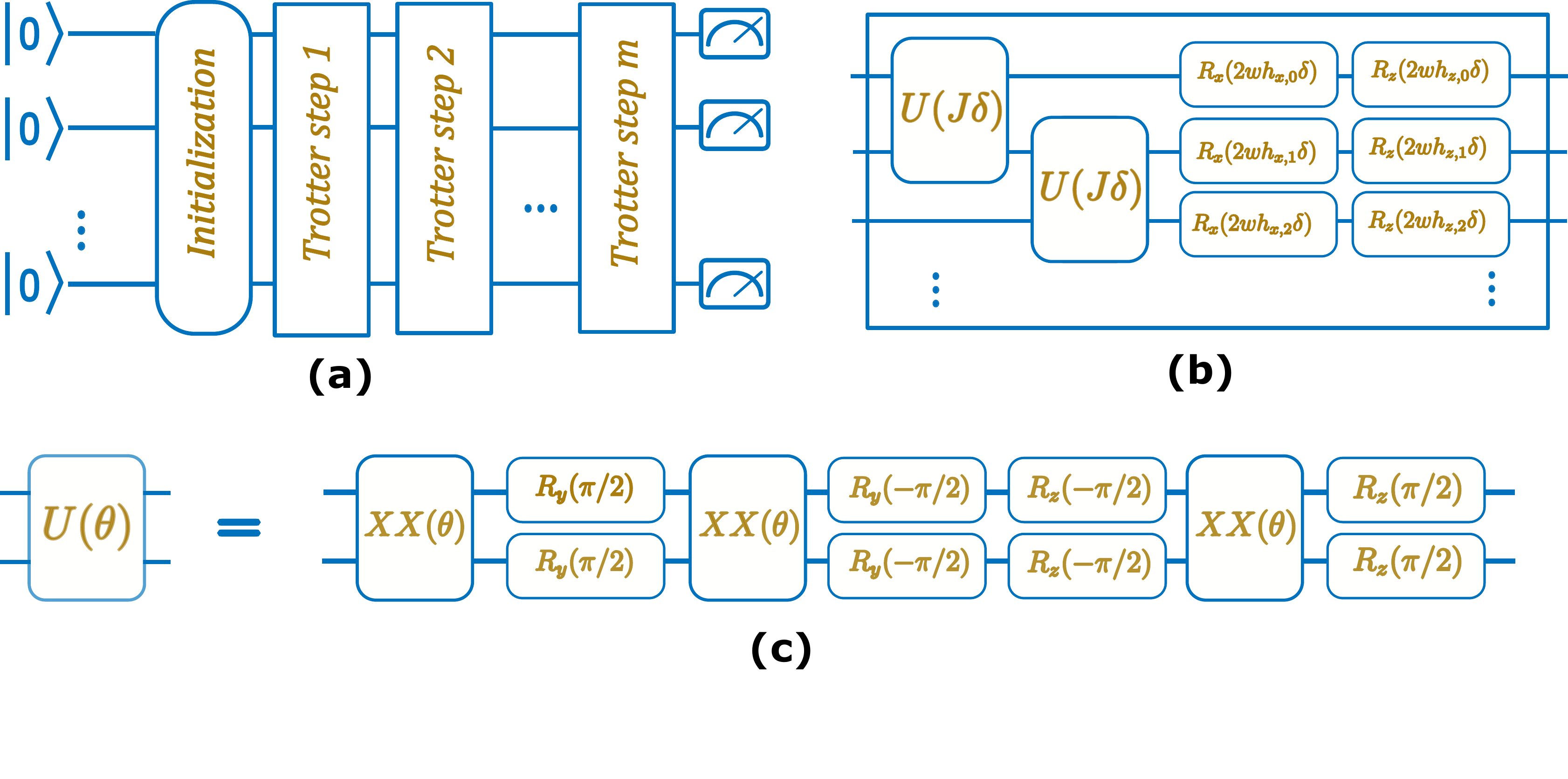}
 \caption{The circuit used to simulate time-evolution under the Heisenberg model Hamiltonian. (a) After the qubits are prepared into the desired initial state, $m=6$ Trotter steps are used to evolve the system to time $t$. All qubits are then measured in the $z$-basis. (b) Each Trotter step consists of several one- and two-qubit gates, as described by Eq. \ref{eq:U}. The single-body interactions are implemented as rotations about the $X$ or $Z$ axis ($R_x$ and $R_z$ gates). (c) The two-body interactions $U(J\delta)$ are implemented as three XX(Ising) gates sandwiched between single-qubit rotations. This segment of the circuit is equivalent to a sequential application of XX, YY, ZZ gates, which describes evolution under the Heisenberg interaction exactly.}\label{fig:circuit}
\end{figure*}\label{fiand g:circuit}

 %\section{Theory}
For a given Hamiltonian $H$ with eigenstates $\ket{\phi_m}$ and corresponding eigenenergies $E_m$, the spectral function of an operator $\hat{a}$ is defined as
\begin{align}\label{eq:spectral_fn}
    A(\omega)=\sum_{l,k} |\langle \phi_l|\hat{a}|\phi_k\rangle| \delta_{\omega,E_k-E_l},
\end{align}
%It is easy to see that $A(\omega)=A(-\omega)$. For $\omega\neq 0$, $A(\omega)$ provides insight into the properties of the off-diagonal elements of the operator in the eigenstate basis.

where we take $\hbar=1$.

For our study, we choose the one-dimensional Heisenberg model with random fields along two axes which for $n$ spins has the Hamiltonian
\begin{align}
    H=J\sum_{i=1}^{n-1} \vec{\sigma}_i. \vec{\sigma}_{i+1} +w\bigg(\sum_{i=1}^n h_i^x \sigma_i^x + \sum_{i=1}^n h_i^z \sigma_i^z \bigg).
\end{align}
Here, $\vec{\sigma}_i=(\sigma_i^x,\sigma_i^y,\sigma_i^z)$ are the Pauli operators. The disorder in the model comes from the fields $h_i^x$ and $h_i^z$, which are random variables chosen from a uniform probability distribution between -1 and 1. In the limit $w/J\to 0$, the system is in the thermalized phase and for $w/J\to\infty$, it is in the localized phase. This model is known to have a phase transition at $w/J \sim 6$ \cite{Geraedts_2017}. We set $w = 1$. 
%The Heisenberg coupling has not been implemented in a quantum simulator yet, but can be straightforwardly implemented on a universal quantum computer.

For a local operator $\hat{a}$ such as a single spin Pauli operator, the spectral function $A(\omega)$ for $n$ spins at $J=0$ will consist of $2n$ delta functions at $\pm 2w\sqrt{{h_i^x}^2+{h_i^z}^2}$. The average spacing between the peaks is $\sim w/n$. For $0<J<<w$, each peak of the non-interacting spectrum will split into a cluster of delta functions with a hierarchy of energy gaps  \cite{johri_mbl, nandkishore}. The full width of the cluster is $J\exp(-1/\xi)$, where $\xi$ is the localization length which is an increasing function of $J/w$ (See Appendix \ref{app:linewidths}). When the system is coupled to a thermal bath, the spectral lines broaden and the discrete structure gradually vanishes as the coupling strength increases. It disappears only when the coupling becomes comparable to $J$. In contrast, in the thermal phase, $A(\omega)$ is expected to become an increasingly smooth function of energy as $n$ increases. Here we construct the probability distribution of the widths of these clusters from the linewidths $\Gamma$ of the peaks in the spectrum.

%This is contrast to the level statistics measured in \cite{martinis1}, for which the critical coupling at which the localization signature is lost is exponentially small in the size of the bath. 

%In finite-sized samples, while the spectrum will not be completely smooth in the thermal phase, the linewidths $\Gamma$ are expected to be wider on average than in the localized phase. 

After averaging over spin locations and disorder realizations,  the ratio of the averaged spectrum of the localized phase to that of the thermalized phase should go to zero as $\omega\rightarrow 0$ \cite{nandkishore}. This implies that in the localized phase, local operators are less likely to connect nearby energy eigenstates, instead mixing them and giving rise to level repulsion. The width of the resulting spectral soft-gap is a function of $w$ and remains finite in the thermodynamic  limit.  In contrast, in the thermalized phase, the spectral function decays as $\omega$ increases for $\omega<J$ \cite{mbl_review_1}.

%In contrast, in the thermalized phase, the energy repulsion is proportional to the many-body level spacing which is exponentially small in the system size. 
%Therefore the probability of a transfer between eigenstates caused by a local operator is suppressed as the energy of the transfer goes to zero.
%$\overline{A}(\omega)$ in any many-body localized system will have a soft gap near zero frequency compared to the thermalized phase.  To see this, consider a basis in which the operator $\hat{a}$ is diagonal. In this basis, it is easy to see that, at $J=0$, $\overline{A}(\omega)$ is a constant function arising from the flipping of individual, non-interacting spins. As $J$ is turned on, states that are closer in energy will mix more than states that are distant in energy, leading to larger values of $\bra{\phi}\hat{a}\ket{\phi}$.

As we now show, the spectral functions can be approximated on a quantum computer by Hamiltonian time evolution, followed by measurement of the expectation value of the local operator and a Fourier transform of the resulting time series data. At $t = 0$, let the system be in the state
\begin{align}
    |\Psi(t=0)\rangle=\sum_k c_k |\phi_k\rangle,
\end{align}
where $|k\rangle$ are the eigenstates of the system. The expectation value of operator $\hat{a}$ at time $t$ is
\begin{align}\label{eq:initial_state}
    \langle \hat{a}(t)\rangle=\sum_{k,l} c_k c_l^* a_{kl} e^{-i(E_k-E_l)t},
\end{align}
where $a_{kl}=\langle \phi_l|\hat{a}|\phi_k\rangle$. The absolute value of the Fourier transform of the above expression gives
\begin{align}\label{eq:ft}
    \mathcal{F}\{\langle \hat{a}\rangle\}=\sum_{k,l}|c_k c_l^* a_{kl}| \delta_{\omega,E_k-E_l}.
\end{align}
Note the similarity to the spectrum of $\hat{a}$ from Eq.\ref{eq:spectral_fn}, especially when the initial state (Eq. \ref{eq:initial_state}) is spread over the eigenstates of the system.
%For our first analysis where we examine $A(\omega)$ averaged over disorder distributions, we can expect it to be a good approximation of $A_0(\omega)$. \red{not clear why these two sentence are set in contrast, what is done in the first case that is not done in the second?} Later when we calculate the average discreteness of spectra of individual samples, Eq. \ref{eq:ft} suffices since it differs from Eq. \ref{eq:spectral_fn} only in the heights of the delta functions and not in the locations. 
In the experiment we use $\hat{a} = \sigma_i^z$, and initialize the qubits in the $\ket{+}$ state, which is an equal superposition of the two eigenstates of $\hat{a}$. We measure in the $z$ basis at the end, in order to extract the spectral function corresponding to $\sigma^z_i$ for qubit $i$. When discussing the experimental measurement of $\overline{A}(\omega)$, we are referring to the expression in Eq. \ref{eq:ft} after disorder-averaging.

%In the thermal phase, when each eigenstate is delocalized over the entire Hilbert space, the $c_m$ are all of similar magnitude for a generic starting state and Eq. \ref{eq:ft} approximates $A_0(\omega)$. 

%In the localized phase, the eigenstates are perturbed around product states of spins parallel (or anti-parallel) to the on-site fields which point in a random direction in the $XZ$ plane. Here, we initialize the qubits in the $|+\rangle$ state which should well approximate an equally weighted sum over pairs of eigenstates in Eq. \ref{eq:ft}. We measure the qubits in the $z$ basis at the end, in order to extract the spectral function corresponding to $\sigma^z_i$ for qubit $i$. 

%and measured in the $z$ basis at the end.  Eq. \ref{eq:ft} will approximately give $A_0(\omega)$ in the localized phase as well.

%For a generic starting state, the $c_m$ are of similar magnitude and Eq. \ref{eq:ft} will approximately give $A_0(\omega)$. We find that picking a starting state that is not an eigenstate of $\hat{a}$ gives good results. In our experiment we use $\hat{a} = \sigma_i^z$. The qubits are initialized in the $|+\rangle$ state and measured in the $z$ basis at the end, in order to extract the spectral function corresponding to $\sigma^z_i$ for qubit $i$. 
%We note that the results will hold generically for the spectral functions should hold for any local operator that can be measured. 
%Note that even if the $c_m$ are not uniformaly distributed, the expressions in Eq. \ref{eq:spectral_fn} and \ref{eq:ft} will differ in the heights of the delta functions and not in the locations.

The experiment is conducted on a re-configurable system with up to 9 qubits, where each qubit is realized by the hyperfine-split ground states of a ${}^{171}\textrm{Yb}^+$ ion. Here we use $n=3$ qubits. A universal set of quantum gates consisting of arbitrary single-qubit rotations and XX(Ising)-gates between any pair of qubits can be applied (see the Appendix for details). The study of disordered systems requires averaging over many disorder realizations. We run the experiment for different values of the coupling strength: J=0.1, J=0.3 and J=0.7, using 24 circuits each to sample instances of disorder. The circuits are generated beforehand and fed to the experiment control computer in batches. As long as the ions stay trapped, the system automatically executes the circuits sequentially. Because of a slow drift in the laser alignment, the individual and two qubit rotation angles change over time. We compensate for this by running an automated re-calibration routine after the execution of every three circuits.

The quantum circuit corresponding to the time evolution under $H$ is shown in Fig.\ref{fig:circuit}. The Hamiltonian evolution cannot be implemented exactly and we use Trotterization to decompose it into one and two-qubit gates (Fig. \ref{fig:circuit} (b)). The two-qubit interaction is exactly captured by the unitary $U$ since $XX$, $YY$ and $ZZ$ terms commute with each other. Each Trotter step achieves the following unitary:
\begin{equation}\label{eq:U}
\begin{split}
  U_H=\prod_{k=1}^n( e^{-ih_{z,k}Z_k\delta}e^{-ih_{x,k}X_k\delta} e^{-iJZ_k Z_{k+1}\delta}\\\times e^{-iJY_k Y_{k+1}\delta}e^{-iJX_k X_{k+1}\delta}).
\end{split}
\end{equation}
The total evolution time is given by $t=m\delta$. It is straightforward to extend the circuit to an arbitrary number of qubits. 

The time evolution is sampled at 10 different equally spaced intervals between  $0< t\leq10$. The expectation value of $\sigma_z$ at $t=0$ is trivially known to be zero. We use a constant number $m=6$ Trotter steps for each sample time making the Trotter angle $\delta=t/6$. This is in contrast to the more widely-used method of Trotterization where $\delta$ stays fixed and the number of Trotter steps increases with time. Since the number of Trotter steps is constant no matter the time, the magnitude of experimental error is the same in every circuit \cite{linke_renyi}. We will see that this becomes critical to the error mitigation technique we introduce below.

Each circuit is measured 2400 times to sufficiently reduce the statistical error. We initialize the qubits into $\ket{+}$ states with Hadamard gates. Because the $\ket{+}$ state is an eigenstate of the $U$ operator, we can skip the application of the first set of $U$ gates on all qubits. Each circuit thus consists of 30 two-qubit gates and 116 single-qubit gates. We run a total of 792 circuits to obtain the data in this paper.

A discrete Fourier transform is then applied to the time series for each instance to obtain the spectrum. In the thermodynamic limit, $J=0.1$ lies in the localized phase, $J=0.7$ in the thermalized and $J=0.3$ near the phase transition. For a small system, there is no sharp phase transition but we expect to see a change from thermalized to localized behavior as we lower the value of $J$.

%\section{Experiment}
\begin{figure}
\centering
 \includegraphics[width=\columnwidth]{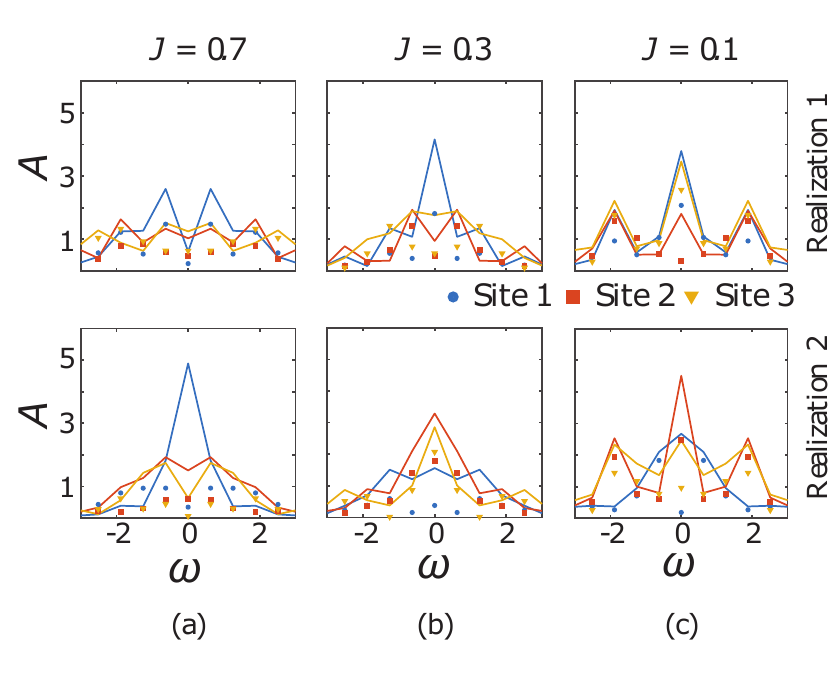}
 \caption{The spectrum of $\sigma^z_i$ at different values of J (with $w=1$) for a 3 site system for two sample disorder realizations (top and bottom). Each panel show both simulation (curves) and experimental (symbols) results. The different colors are for the different sites. The lack of distinguishing characteristics between the spectra at different values of $J$ for individual samples shows the necessity of averaging over several disorder realizations.}\label{fig:realizations}
\end{figure}

Fig. \ref{fig:realizations} shows several instances of the measured spectrum for $\sigma_i^z$. The spectrum is symmetric about $\omega=0$. We note that the experimental data is significantly damped compared to the simulation. The figure also shows the necessity of averaging over several realizations in the study of disordered systems since the behavior of the system in the thermodynamic limit cannot be determined from the behavior of a finite-size individual disorder realization.

\begin{figure}
\centering
 \includegraphics[width=\columnwidth]{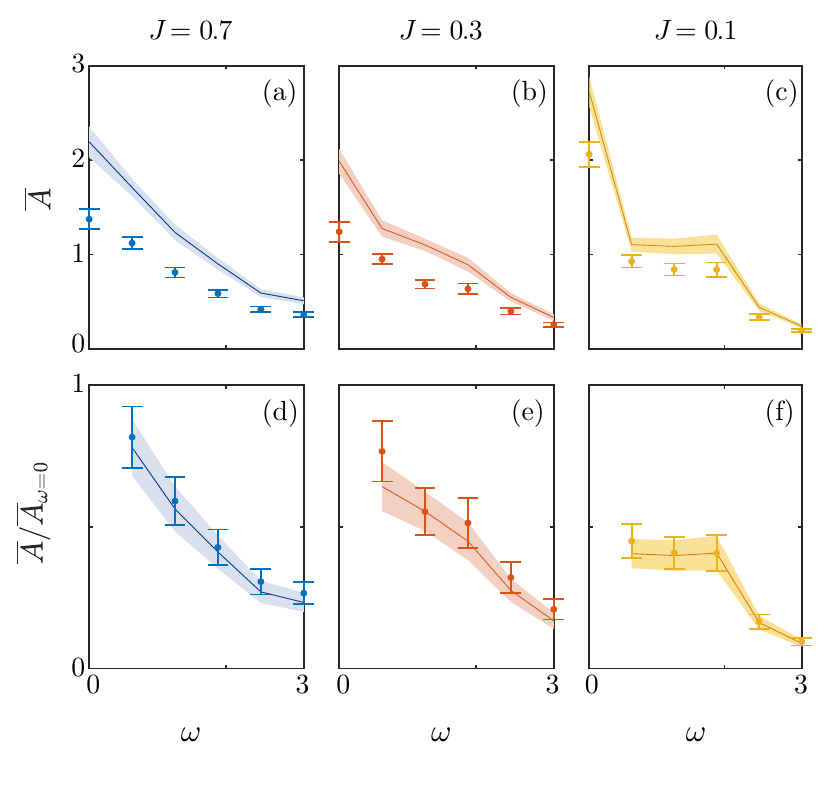}
 \caption{(a)-(c) The spectral function of $\sigma^z_i$ averaged over position and 24 disorder realizations for different values of $J$ (with $w=1$) for a 3-site system. Lines show simulation results and circles show experimental results. (d)-(f) The spectral function normalized by its value at $\omega=0$. At $J=0.7$, the spectral response increases as $\omega$ decreases while at $J=0.1$, the spectrum is damped at low frequencies. The points at $\omega=0$ in the top row are discontinuous with the rest of the curves since they arise from the diagonal elements of the observable in the eigenstate basis which have qualitatively different behavior than the off-diagonal ones. %This indicates the suppression of transitions between eigenstates mediated by the local operators $\sigma_i^z$ due to localization when disorder dominates interactions.
 }\label{fig:avg_spectrum}
\end{figure}

We next average the spectral functions over lattice sites and disorder configurations. The results are shown in Fig. \ref{fig:avg_spectrum} (a)-(c). The value at $\omega=0$ arises from the diagonal elements and is related to the equilibrium value of the observable at infinite temperature, whereas the behavior at $\omega>0$ gives the dynamical response of the system. The simulation curves show that the value of the spectral function at low-frequencies drops as $J$ decreases. Simulation results for a larger system size show similar behavior (Appendix B). While the experimental data follow the trend of the simulation for each value of $J$, the error obscures the difference between the spectra at different values of $J$. To address this, we now introduce an error mitigation technique.

%Here, we expect to see a fall-off in the response at low frequencies in the localized phase, $J=0.1$. This is indeed the case in the simulation results as shown in Fig. \ref{fig:avg_spectrum} (a). We see that while the experiment obeys the general trend of the simulation, it is quite hard to distinguish localized and thermalized behavior.  (Appendix B also shows a simulation of the behavior of the disorder-averaged spectral function when $J$ is fixed and $w$ is varied.)

% 
It has been shown that the error in the mean value of an observable measured after the application of a set of random circuits with the same structure can be well-approximated by a depolarizing error model, whatever the origin of the noise \cite{google_qs}. Therefore, the mean density matrix after the application of the unitaries $\{U_H\}$ in Eq. \ref{eq:U} to an initial state $\ket{\Psi_0}$ is
\begin{align}
    \overline{\rho}=\epsilon_m\overline{U_H\ket{\Psi_0}\bra{\Psi_0}U_H^{\dagger}}+(1-\epsilon_m)\frac{I}{D},
\end{align}
where $I$ is the identity matrix and $D=2^n$. $\epsilon_m=p^m$, where $p$ is the disorder-averaged depolarization fidelity per Trotter step. The expectation value of $\hat{a}$ at time $t$ is
\begin{align}
    \overline{\langle \hat{a}\rangle} (t)=\Tr(\overline{\rho}(t)) \hat{a})=p^m\Tr(\overline{U_H(t)\ket{\Psi_0}\bra{\Psi_0}U_H^{\dagger}(t)}\hat{a}).
\end{align}
Since the same number of Trotter steps $m$ is used for measuring at all times, the corresponding spectrum obtained by Fourier transform becomes
\begin{align}
     \overline{A}(\omega)=p_c^m \int \Tr(\overline{U_H(t)\ket{\Psi_0}\bra{\Psi_0}U_H^{\dagger}(t)}\hat{a}) e^{-i\omega t}dt.
\end{align}

If we now divide by the zero-frequency component,
\begin{align}
     \frac{\overline{A}(\omega)}{\overline{A}(0)}=\frac{\int \Tr(\overline{U_H(t)\ket{\Psi_0}\bra{\Psi_0}U_H^{\dagger}(t)}\hat{a}) e^{-i\omega t}dt}{\int \Tr(\overline{U_H(t)\ket{\Psi_0}\bra{\Psi_0}U_H^{\dagger}(t)}\hat{a}) dt},
\end{align}
$p^m$ is canceled. We should thus essentially get a noiseless signal after the normalization. Fig. \ref{fig:avg_spectrum} (d)-(f) shows the normalized spectra. We see that the data match the normalized curves within the statistical uncertainty, especially at $J=0.1$ and $J=0.7$ which are deep in the localized and thermalized phase respectively.  Note that the estimated fidelity of the quantum computation obtained by multiplying the fidelities of the individual gates is only 54\%, making the experimental reproduction of the theoretical curves in Fig. \ref{fig:avg_spectrum}(b) remarkable. %Normalization also seems to magnify the difference between the localized and thermalized behavior in the ideal simulated curves.%, though this effect needs to be verified by simulation of larger system sizes.
%This is equivalent to normalizing the curve by the DC response. 

%The normalized spectrum at $J=0.3$ which lies near the thermodynamic phase transition has the most error.

%Note that is the style of Trotterization, i.e. choosing the same number of Trotter steps per measurement time that made this correction possible.

\begin{figure}
\centering
 \includegraphics[width=0.95\columnwidth]{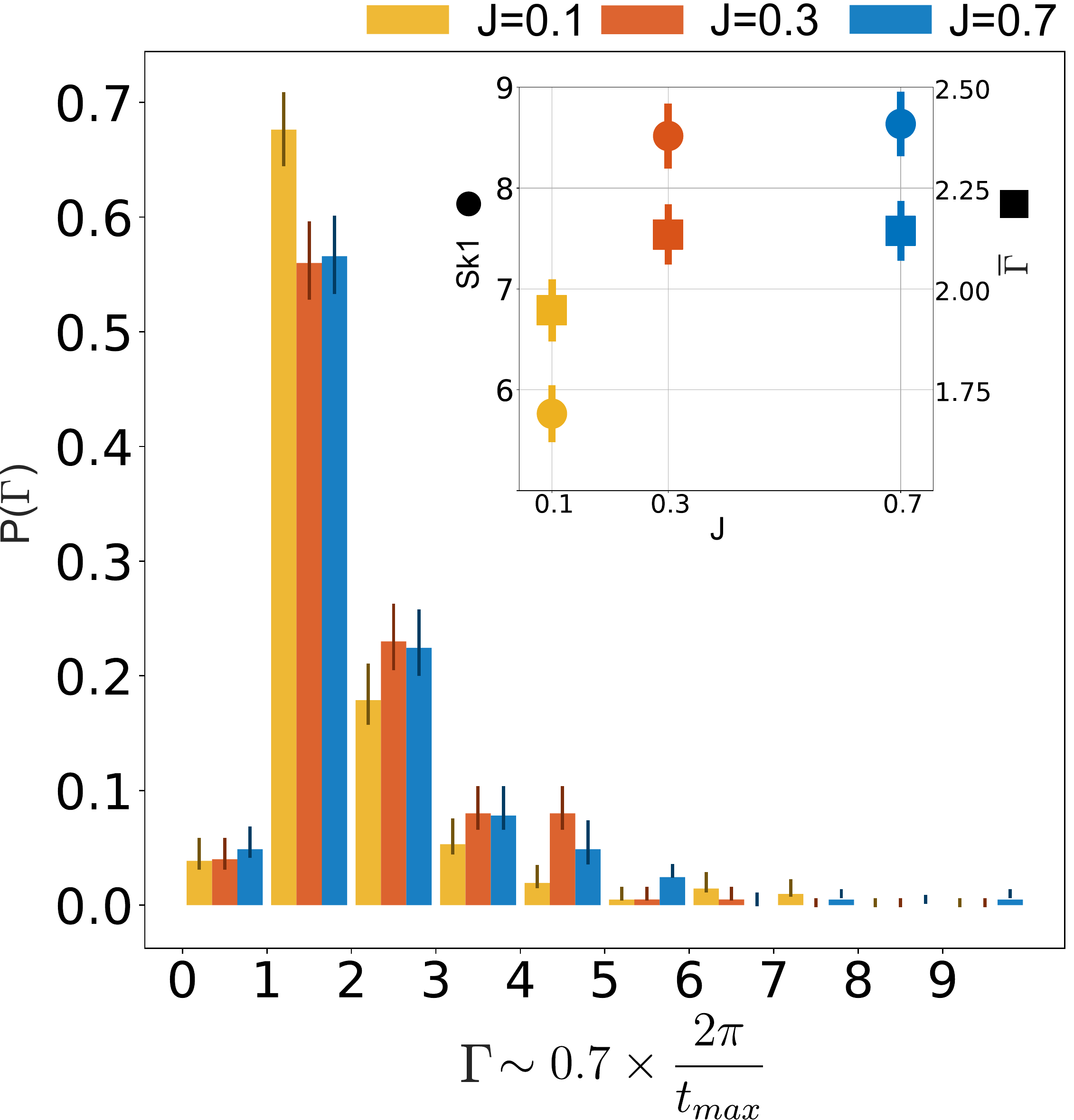}
 \caption{The distribution histogram of linewidths $\Gamma$ calculated as described in the text at different values of $J$ from data taken on 3 qubits for 24 realizations.The bins are [0-1], [1-2], ... etc. The number of peaks used to generate the distribution is $\sim 200$ for each value of $J$. We derive the errorbars shown in the plot by assuming each bin approximately follows a binomial distribution.  The inset shows the Pearson's first coefficient of skewness, Sk1, and the average linewidth, $\bar{\Gamma}$. %he points correspond to the height of the histograms obtained as described in the text and the lines are smoothing fits. 
 }\label{fig:line_widths}
\end{figure}

We next test the discreteness of the distribution by studying the linewidths $\Gamma$ of the peaks in the spectrum. We expect the peaks in the localized phase to be narrower than in the thermalized phase on average. As shown in the Appendix of Ref \cite{johri_mbl}, the distribution should be be skewed to larger linewidths, indicating the presence of resonant clusters of spins. 

We use the following procedure to obtain the probability distribution of the linewidths:

1. Fit individual spectra (such as those in Fig. \ref{fig:realizations}) with an interpolating polynomial and find the peaks.

2. For each peak, find the width corresponding to half the prominence of the peak.

3. Plot a normalized histogram corresponding to the probability distribution of the line-widths thus obtained.

The probability distributions $P(\Gamma)$ are shown in Fig. \ref{fig:line_widths} for different values of $J$. As expected, they are skewed to the right. In the inset of Fig. \ref{fig:line_widths}, we show that both the average linewidth as well as the skewness, which measures the probability of resonant clusters, are smaller at $J=0.1$ than at larger values of $J$. See Appendix C for a more detailed picture of how the spectrum changes with $J$.
%\section{Conclusion}

%\red{I suggest suppressing this paragraph}

In conclusion, we have presented the first study of spectral functions of local operators that carry noise-resilient signatures of localization on a quantum computer. Since spectral functions determine transport properties, we anticipate that this algorithm along with the corresponding error mitigation technique will be useful in materials design applications of quantum computers.
%Spectral functions carry signatures of localization such as discreteness and a universal soft gap at low frequencies that are inherently robust to noise. Additionally, we develop an error mitigation method that can extract useful information even when the signal-to-noise ratio is very low. We have demonstrated the technique for a disordered Heisenberg model on trapped ion qubits after averaging over 100s of circuits. 

%In this study, we chose a low number of qubits and Trotter steps based on expectations of the fidelity of the circuit from the fidelity of the individual gates. However, 
Our error mitigation has worked so well that we foresee that the circuits run here could be extended to many more qubits and gates without a significant loss in the quality of the results. We hope to show this in future work. In particular, it is promising that the error mitigation used here did not require any extra data from the quantum computer. We encourage researchers to develop similarly scalable and noise-resistant techniques for studying other unsolved problems in condensed matter physics on quantum computers.

\section*{Acknowledgements}
This work was supported by the ARO with funds from the Intelligence Advanced Research Projects Activity (IARPA) LogiQ program (Grant Number W911NF16-1-0082), the Army Research Office (ARO) MURI program on Modular Quantum Circuits (Grant Number W911NF1610349), the AFOSR MURI program on Optimal Quantum Measurements (Grant Number 5710003628), the NSF STAQ Practical Fully-Connected Quantum Computer Project, and the DOE BES Quantum Computing Program (Grant Number de-sc0019449).  C.H.A. acknowledges financial support from CONACYT doctoral grant No. 455378.

\pagebreak

\appendix

\section{Experimental Details}

The system is based on a chain of ${}^{171}\textrm{Yb}^+$ ions held in an RF Paul trap \cite{debnath2016demonstration}. Each ion provides one physical qubit in the form of a pair of states in the hyperfine-split ${}^{2}\textrm{S}_{1/2}$ ground level with an energy difference of 12.642821 GHz, which is insensitive to magnetic field fluctuations to first order. The qubits are initialized to $\ket{0}$ by optical pumping and read out by state-dependent fluorescence detection \cite{Olmschenk07}. Gates are realized by a pair of Raman beams derived from a single 355-nm mode-locked laser. These optical controllers consist of a global beam that illuminates the entire chain and an array of individual addressing beams. Single-qubit rotations are realized by driving resonant Rabi rotations of defined phase, duration, and amplitude. Two-qubit gates are achieved by illuminating two selected ions with beat-note frequencies near the motional sidebands creating an effective Ising spin-spin interaction via transient entanglement between the two qubits and the motion in the trap \cite{Molmer99,Solano99,Milburn00}. Our scheme involves multiple modes of motion, which are disentangled from the qubits at the end of an two-qubit gate operations via an amplitude modulation scheme\cite{choi2014optimal}. The effect of this scheme can be described by the unitary $\exp(i \hat{X}_i \hat{X}_j \chi )$, where $\hat{X}_i$ stands for the Pauli X operator of qubit $i$. This type of gates are known as XX or Ising gate. Typical single- and two-qubit gate fidelities are $99.5(2)\%$ and $98-99\%$, respectively. The latter is  limited by residual entanglement of the qubit states and the motional state of the ions due to intensity noise, and motional heating. Rotations around the z-axis are achieved by phase advances on the classical control signals.

\section{Simulation for larger system sizes}
\begin{figure}
\centering
 \includegraphics[width=\columnwidth]{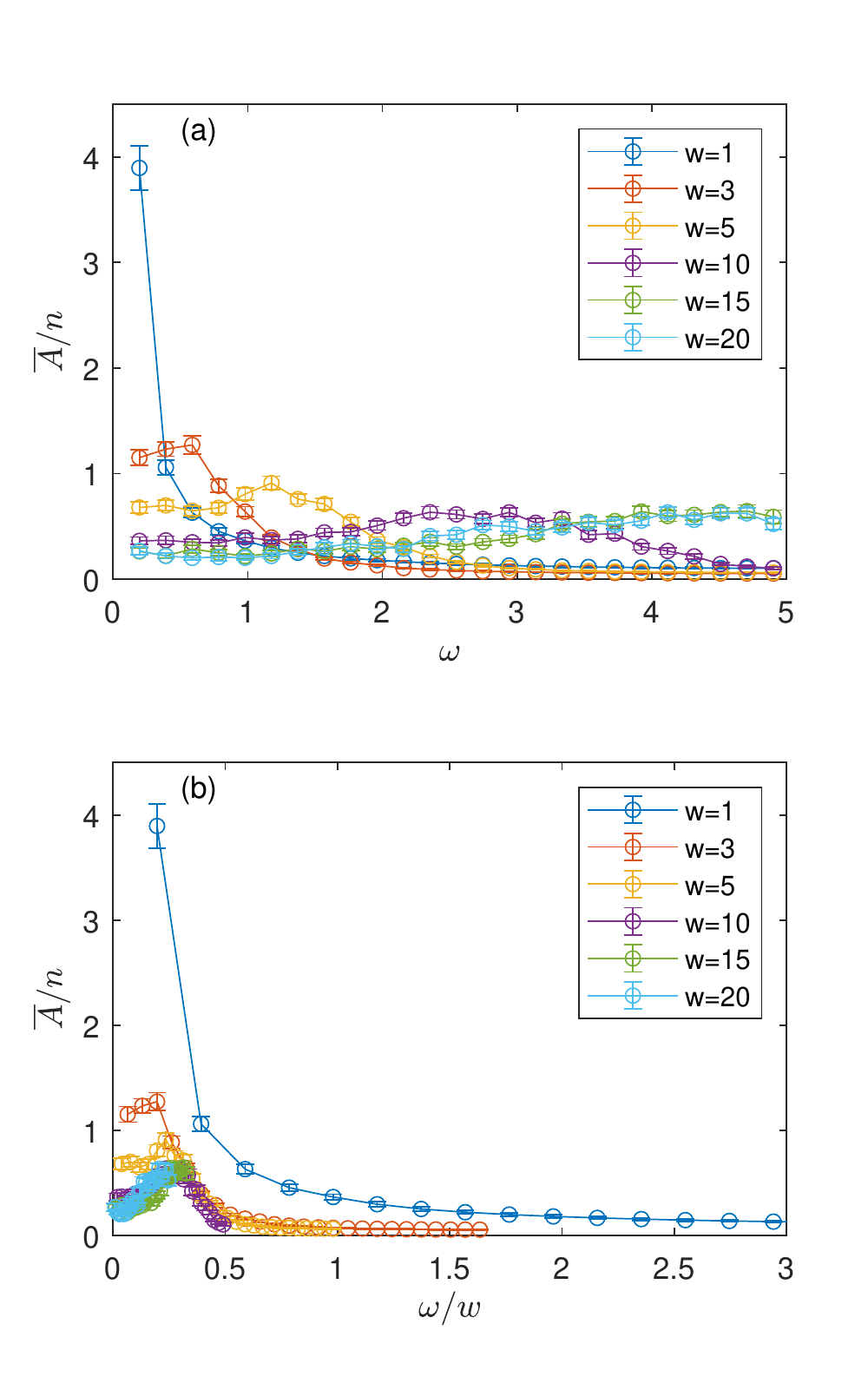}
 \caption{(a) The simulated spectrum for $n=7$ spins averaged over 100 disorder configurations for fixed $J=1$, while $w$ is varied. (b) The same plot with the frequency scaled by $w$.}\label{fig:spectrum_vary_w}
\end{figure}

Fig. \ref{fig:spectrum_vary_w} (a) shows simulations for a 7 spin system for which $J=1$ and $w$ is varied using 500 Trotter steps for each sample time and averaging over 100 disorder realizations. As $w$ increases, the maximum of the spectrum shifts right while its magnitude at low frequencies goes down. The high-frequency regime represents one-body physics for which the energy scale is set by $w$. Therefore, in Fig. \ref{fig:spectrum_vary_w} (b) which shows the same spectra on a plot where the frequency $\omega$ has been scaled by the disorder magnitude $w$, the curves now lie on top of each other at high-frequencies. Note that this plot is equivalent to fixing $w=1$ while varying $J$, and plotting the spectrum versus $\omega$ as is done for the data in Fig. \ref{fig:avg_spectrum} of the paper. At low frequencies, there is a suppression of the spectral function as the ratio $w/J$ increases, consistent with the results presented in the main text for a 3 spin system.

\section{Linewidths}\label{app:linewidths}
\begin{figure}
\vspace{10pt}
\centering
 \includegraphics[width=0.8\columnwidth]{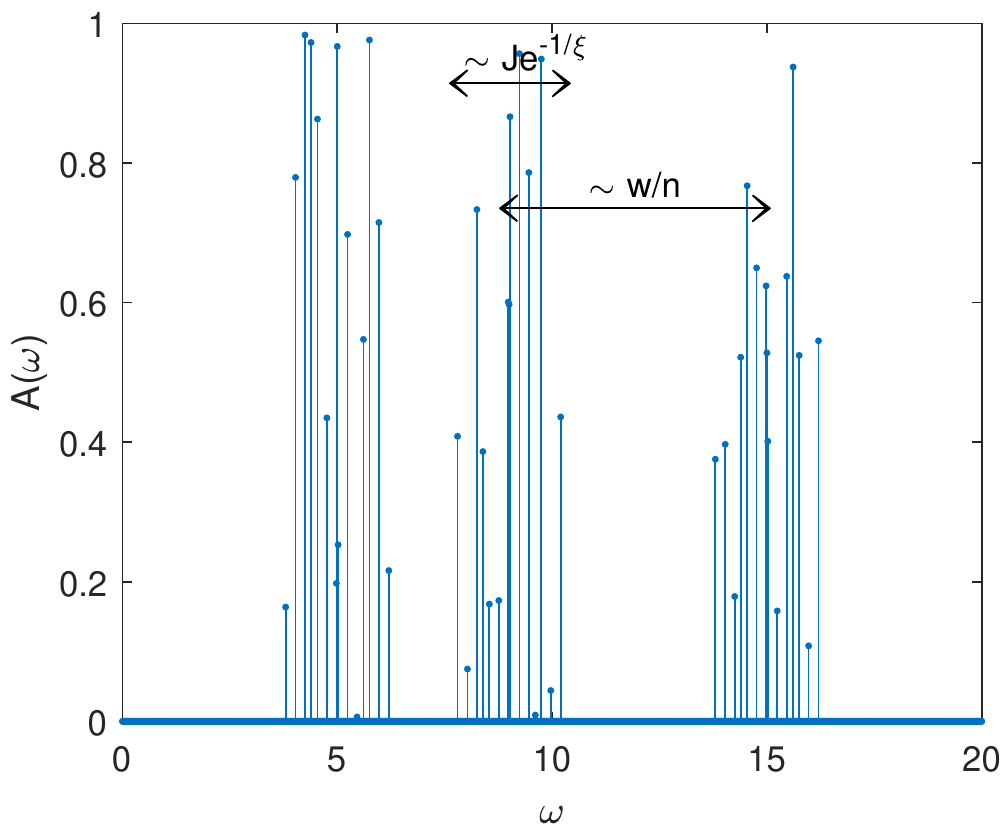}
 \caption{A schematic example of the splitting of spectral lines according to the theoretical model in \cite{nandkishore} for a localized system in the limit $J<<w$.}\label{fig:sample_spectrum}
\end{figure}

When $J=0$, the spectrum consists of $n$ delta functions at $\omega=\pm 2w\sqrt{{h^x_i}^2+{h^z_i}^2}$. When the interaction $J$ is turned on, each delta function branches into a tree like structure. The splitting at each branch of the tree is proportional to $J\exp(-d/\xi)$, where $d$ is the depth of the branch and $\xi$ is the localization length \cite{nandkishore}. Fig. \ref{fig:sample_spectrum} shows a schematic example of this. When $J$ is large enough, no discrete structure will remain and the spectrum will be continuous, indicating a transition to thermalization. In our experiment, we do not sample at enough points in the time-evolution to resolve the hierarchical structure of the energy gaps but we can measure the total broadening of the original spectral line. The results for this are presented in Fig. \ref{fig:line_widths} of the main text.

%\begin{figure}
%\vspace{10pt}
%\centering
% \includegraphics[width=0.8\columnwidth]{average_linewidth.pdf}
% \caption{}\label{fig:average_linewidth}
%\end{figure}

%If the Hamiltonian has no disorder, the energy spectrum is highly degenerate. This in turn implies that $A(\omega)$ has several delta functions at $E_i-E_j$. At small disorder, the energy degeneracies will split. However, since the states are still extended, they will be connected by local operators. Therefore, at small disorder, the delta function peaks will broaden, with the line-width proportional to $\sim w^2/J$.

%Contrast this with the case of large disorder. This is discussed in Ref. \cite{nandkishore}. When there are no interactions, there are discrete delta functions of weight 1 in the spectral function. As $J$ increases, these delta functions rearrange themselves as . However, they still remain discrete.

%Therefore, the linewidths in the thermal phase are wider than in the localized phase.

\pagebreak
%\bibliography{reference}


\end{document}